\documentclass[onecolumn,bibyear]{aa} 
\usepackage{graphicx}
\usepackage{txfonts}
\begin{document} 
   \title{The emergence of the galactic stellar mass function from a non-universal IMF in clusters}

   \author{Sami Dib \inst{1,2}, Shantanu Basu\inst{3}}

   \institute{Niels Bohr International Academy, Niels Bohr Institute, Blegdamsvej 17, DK-2100, Copenhagen, Denmark\\
                   \email{sami.dib@gmail.com}
          \and    
                   Department of Physics and Astronomy, University of Western Ontario, London, Ontario, N6A 3K7, Canada}

 
 \abstract{We investigate how a single generation galactic mass function (SGMF) depends on the existence of variations in the initial stellar mass functions (IMF) of stellar clusters. We show that cluster-to-cluster variations of the IMF lead to a multicomponent SGMF where each component in a given mass range can be described by a distinct power-law function. We also show that a dispersion of $\approx 0.3$ M$_{\odot}$ in the characteristic mass of the IMF, as observed for young Galactic clusters, leads to a low mass slope of the SGMF that matches the observed Galactic stellar mass function even when the IMFs in the low mass end of individual clusters are much steeper.}
 

   \maketitle
%

\section{Introduction}

The initial mass function (IMF) of stars (i.e., the distribution of the masses of stars at their birth), is of fundamental importance in astrophysics. The IMF controls the efficiency of star formation in molecular clouds (Nakamura \& Li 2007; Dib et al. 2011,2013; Hony et al. 2015), the radiative and mechanical feedback from stars into the interstellar medium (e.g., Dib et al. 2006; Dib 2011; Padoan et al. 2016; Martizzi et al. 2016) and the dynamical and chemical evolution of galaxies (Bekki \& Tsujimoto 2014). Thus, the characterisation of the correct shape of the mass function of stars both on cluster and galactic scales is of vital importance. For the nearby Galactic field stars in the mass range $0.4 \lesssim M_{\star}/M_{\odot} \lesssim 10$, Salpeter (1955) found that the stellar mass function is well described by a power law $dN/d{\rm log}M_{*}=M_{*}^{-\Gamma}$ (with $\Gamma \approx 1.35$), where $dN$ is the number of stars between ${\rm log}M_{*}$ and ${\rm log} M_{*}+d{\rm log}M_{*}$. Since then, there has been a persistent effort to refine the description of the shape of the IMF in individual stellar clusters and of the stellar mass function in the Milky Way and other galaxies. Various surveys suggest that the mass function of stars in the Galactic field, which is uncorrected for the effects of the binary population, rises from the brown dwarf and low stellar mass regime until it peaks at $\approx 0.25-0.4$ M$_{\odot}$ after which it declines steeply in the intermediate-to-high mass regime (Scalo 1986; Bochanski et al. 2010; Rybizki \& Just 2015). Several distribution functions are used to describe its shape, such as a multi-component power-law (Kroupa 2001; Kroupa et al. 2013), a lognormal coupled to a power-law (Chabrier 2005), a tapered power law (Parravano et al. 2011) or a modified lognormal (Basu et al. 2015).   

  \begin{figure*}
   \centering
    \includegraphics[width=0.45\textwidth]{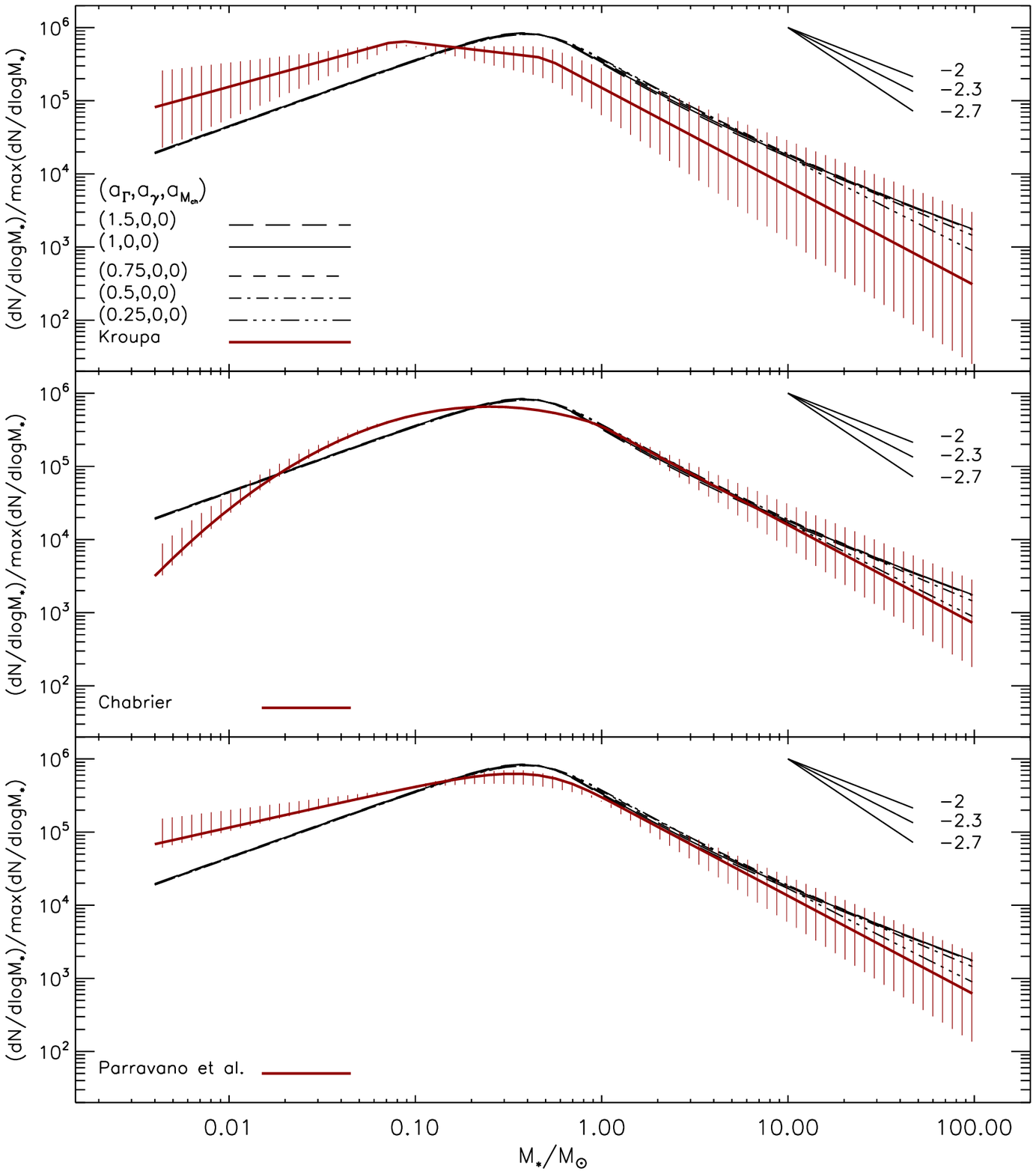}
    \hspace{0.8cm}
    \includegraphics[width=0.45\textwidth]{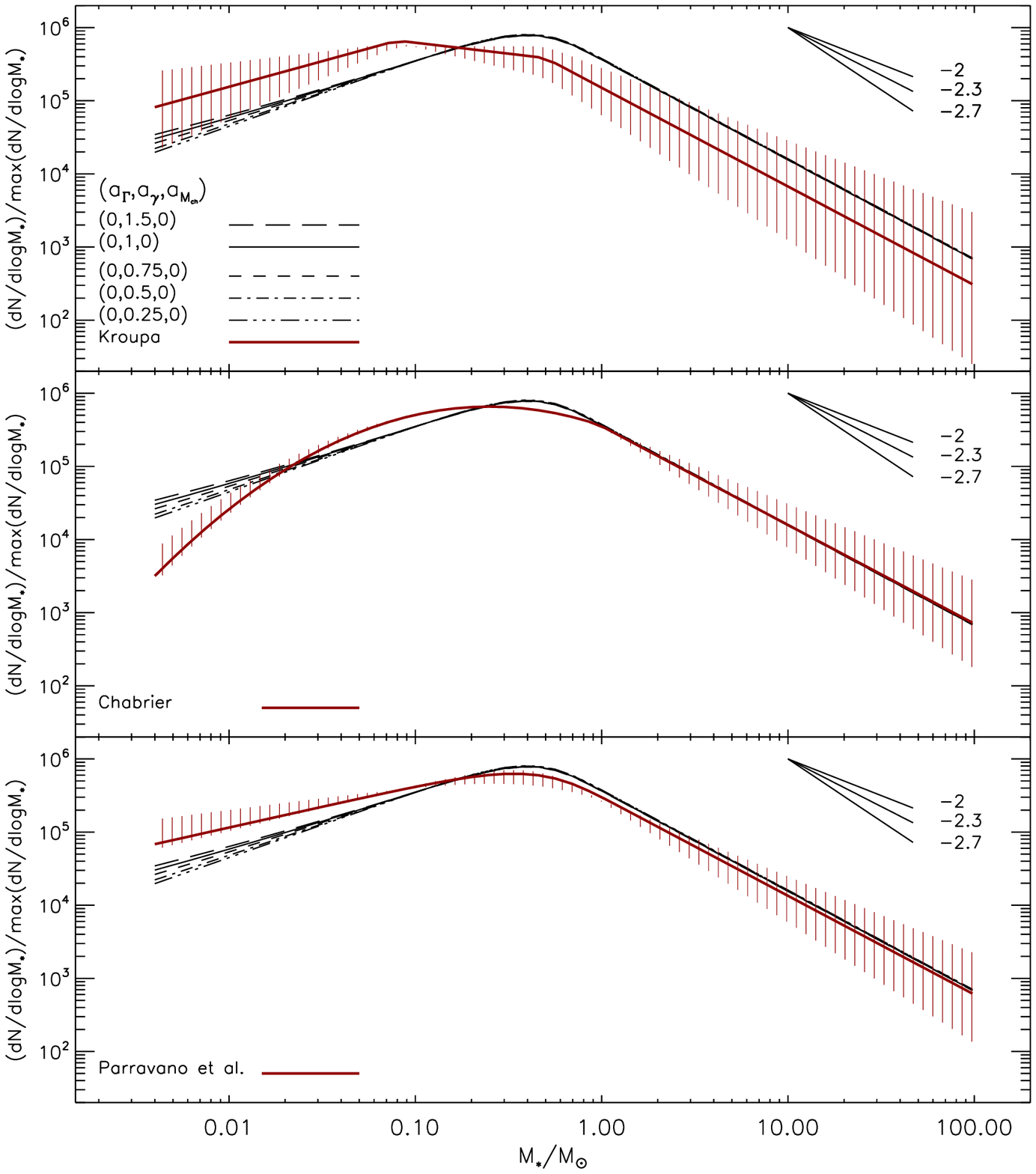}\\
    \vspace{1cm}
    \includegraphics[width=0.45\textwidth]{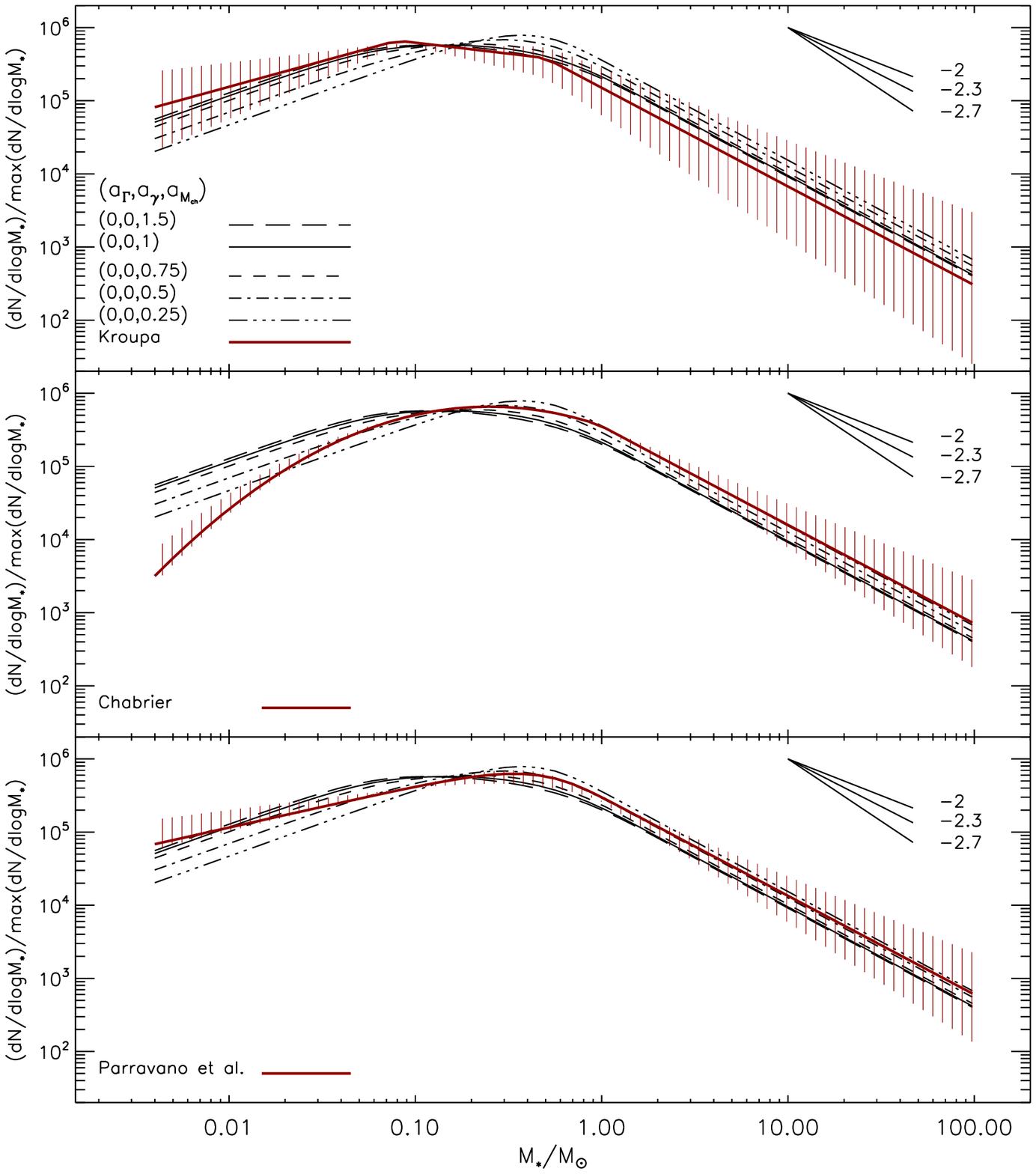}
    \hspace{1cm}
    \includegraphics[width=0.45\textwidth]{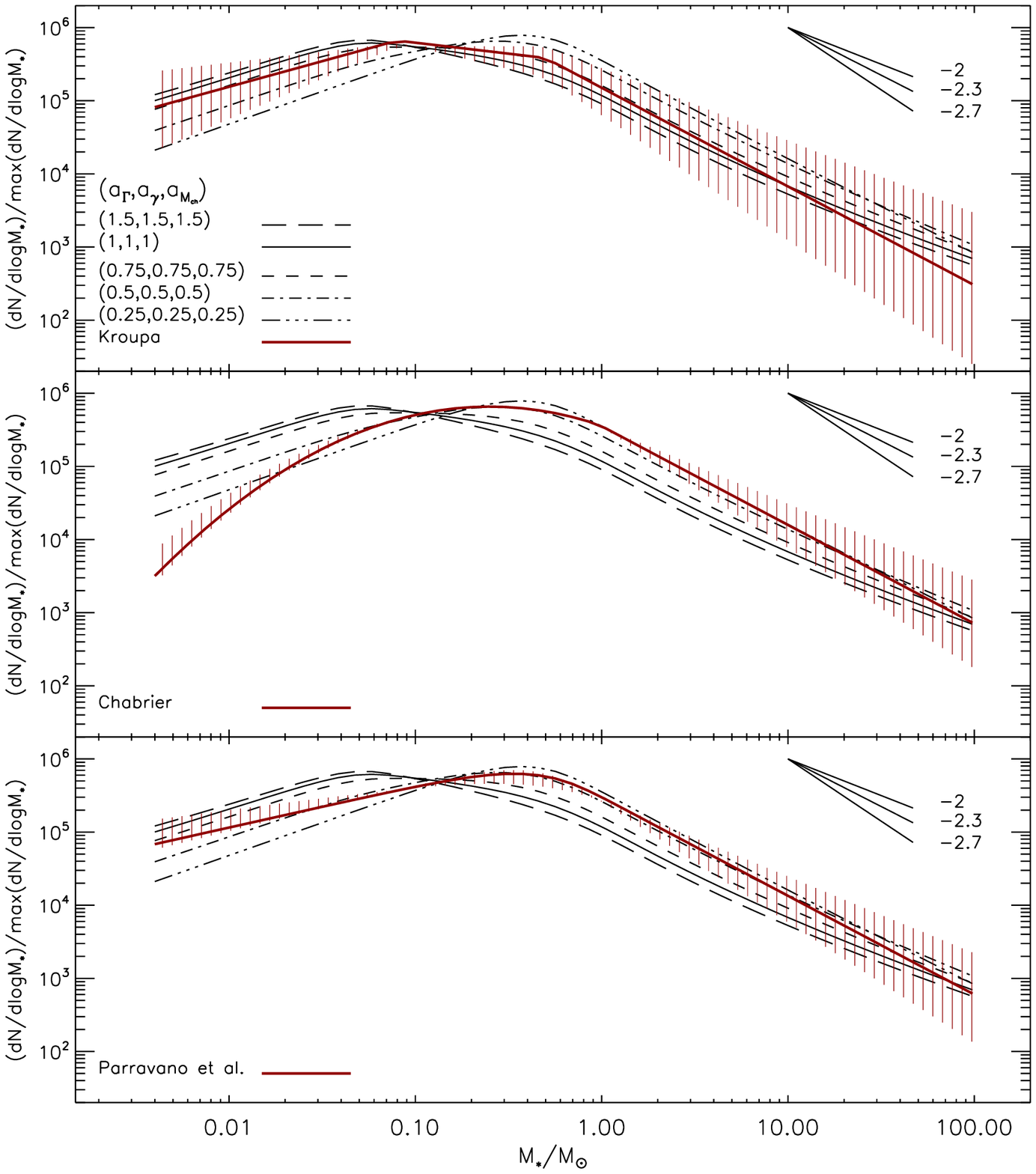}
    \vspace{0.8cm}
      \caption{The single generation Galactic stellar mass (SGMF) for various choices of the dispersion in the three parameters of the stellar clusters IMF: $\Gamma$, $\gamma$, and M$_{ch}$. The three stacked figures in the top left panel show cases with cluster-to-cluster variations only in $\Gamma$, the top right panel displays cases with variations only in $\gamma$, and the lower left panel displays cases that include variations only in M$_{ch}$. The ensemble of black curves in each of the stacked panels are similar. The models in each of  the stacked panels are compared to the Kroupa (2001), Chabrier (2005), and Parravano et al. (2011) Galactic field stellar mass function (red line) with their associated uncertainties (hatched regions in red).}
    \label{fig1}
   \end{figure*}

A large number of studies have also attempted to derive the shape of the IMF\footnote{Through the paper, we strive to keep the terminology as clear as possible. The term IMF is reserved for individual clusters, the term SGMF is the cumulative IMF for a population of clusters, and the term galactic/Galactic mass function refers to the stellar mass function of a galaxy/the Milky Way} of stars in individual stellar clusters. Stars in young clusters (i.e., with ages $\lesssim 10-12$ Myr) that have not undergone an extensive dynamical evolution have roughly the same age and metallicity, and are located at the same distance, thus one can presume that their observed present day mass functions are nearly identical to their IMFs. Some of the most recent studies suggest that there are non-negligible, intrinsic cluster-to-cluster variations in the set of parameters that characterise the shape of the IMF among the population of young stellar clusters in the Milky Way (Dib 2014; Mallick et al. 2014; Dib et al. 2017) but also in globular clusters (Zaritsky et al. 2014). Using a Bayesian approach, Dib (2014) showed that the parameters of the IMF for 8 young Galactic stellar clusters do not overlap at the $1-\sigma$ confidence limit interval. Using a different method, Dib et al. (2017) showed that the fraction of isolated O stars measured in a sample of 341 clusters from the Milky Way Stellar Clusters Survey (MWSC; Kharchenko et al. 2013; Schmeja et al. 2014) can only be reproduced by populations of Galactic clusters that have a substantial intrinsic scatter in the set of parameters that characterise their IMF. 

When connecting the IMF in clusters to the mass function of stars in a galaxy by integrating over an initial cluster mass function (ICMF), all previous studies assume that the IMF of each individual cluster is indistinguishable from that of the mass function of the Galactic field, setting aside its normalisation which is determined by the available stellar mass in each cluster (e.g., Weidner \& Kroupa 2005; Haas \& Anders 2011; Weidner et al. 2013). In this work, we explore how cluster-to-cluster variations of the IMF affect the resulting galactic stellar mass function for a single generation of stars (i.e., the SGMF) and compare the results to the observationally derived Galactic mass function. 
   
\section{THE IMF OF INDIVIDUAL CLUSTERS}\label{individual}
        
We use a description of the IMF in clusters that is given by the tapered-power law function (TPL, de Marchi et al. 2010; Parravano et al. 2011; Dib et al. 2017). This is a convenient form because the TPL can be described with only three free parameters and also because Dib et al. (2017) inferred the appropriate distribution functions of the IMF parameters of young Galactic clusters using a TPL description of the IMF. The TPL function is given by

\begin{equation}
\phi\left({\rm log} M_{*}\right)=\frac{dN_{*}} {d{\rm log}M_{*}}=A_{*}\times M_{*}^{-\Gamma}\left\{1-\exp\left[-\left(\frac{M_{*}}{M_{ch}}\right)^{\gamma+\Gamma}\right] \right\},
\label{eq1}
\end{equation} 

\noindent where $\Gamma$ is the slope in the intermediate- to high mass regime, $\gamma$ the slope in the low mass regime, $M_{ch}$  is the characteristic mass, and $A_{*}$ is the normalisation coefficient which is set by the cluster's mass $M_{cl}$ (i.e., $M_{cl}=A_{*}\int_{M_{*,min}}^{M_{*,max}} \phi({\rm log}M_{*}) dM_{*}$, with $M_{*,min}$ and $M_{*,max}$ being the minimum and maximum stellar masses). Dib (2014) showed that the mean values of $\Gamma$, $\gamma$, and $M_{ch}$ for the star system IMFs (i.e., uncorrected for binarity) for a sample of 8 young Galactic clusters is $\Gamma_{obs}=1.37$, $\gamma_{obs}=0.91$, and $M_{ch,obs}=0.41$ M$_{\odot}$ with standard deviations of $\sigma_{\Gamma_{obs}}=0.6$, $\sigma_{\gamma_{obs}}=0.25$, and $\sigma_{M_{ch,obs}}=0.27$ M$_{\odot}$, respectively. Dib et al. (2017) showed that such a level of intrinsic scatter for each of these parameters is necessary in order to reproduce the fraction of isolated O stars that is measured in young Galactic clusters in the Milky Way. They showed that when the IMF is described by a TPL function, the fraction of isolated O stars measured in a large sample of Galactic young clusters (341 clusters in the MWSC survey) can be reproduced by models in which the parameters of the IMF are drawn from Gaussian functions with with mean values of $\bar{\Gamma}=1.37$, $\bar{\gamma}=0.91$, $\bar{M_{ch}}=0.41$ M$_{\odot}$ , and standard deviations of $\sigma_{\Gamma}=0.6$, $\sigma_{\gamma}=0.25$, and $\sigma_{M_{ch}}=0.27$ M$_{\odot}$, thus matching the direct parameter inference of Dib (2014). Interestingly, three dimensional star-cluster simulations show that such a level of dispersion in the characteristic mass is plausible and can be attributed to variations in the level of turbulent support in star forming molecular clouds (Haugb{\o}lle et al. 2017). Other authors have suggested that variations in all of the IMF parameters can be caused by cloud-to-cloud variations in the mean level of accretion rates onto protostellar cores (e.g., Basu \& Jones 2004, Myers 2009; Dib et al. 2010), or by the coalescence of protostars in high density environments (Shadmehri 2004; Dib et al. 2007; Dib et al. 2008).

\section{A SINGLE GENERATION GALACTIC MASS FUNCTION}\label{sgmf}

For a fixed shape of the IMF, the mass function of a single generation (SGMF) of young stars (i.e., stars in the embedded phase of clusters or over a given timespan in cluster ages $\tau_{cl}$ of the order of $10-12$ Myrs) in a galaxy would be given by

\begin{equation}
\Phi\left({\rm log}M_{*},\tau_{cl}\right)=\frac{d{\cal N}}{d{\rm log}M_{*}}=B_{*}\left(\tau_{cl}\right) \int_{M_{cl,min}}^{M_{cl,max}} \phi\left({\rm log}M_{*},M_{cl}\right) \xi_{cl} \left(M_{cl}\right) dM_{cl},
\label{eq2}
\end{equation}

\noindent where $\xi_{cl}$ represents the ICMF (the mass function of clusters at their birth within the timespan $\tau_{cl}$), $M_{cl,min}$ and $M_{cl,max}$ are the upper and lower mass cut-offs in cluster mass, and $B_{*}$ is a normalization coefficient. The function $\phi\left({\rm log}M_{*},M_{cl}\right)$ is a representation of the IMF (i.e., Eq.~\ref{eq1}) which allows for a potential dependence of the IMF on cluster mass\footnote {The dependence of the IMF on the cluster mass under the assumption of a universal IMF could stem from the existence of a cluster mass-maximum stellar mass relation (Weidner \& Kroupa 2004)}, $M_{cl}$. One common representation of the ICMF is a power-law function with an exponential truncation at the high mass end 

\begin{equation}
\xi_{cl}(M_{cl})=dN/dM_{cl}  \propto e^{-\left(M_{cl}/M_{c}\right)} M_{cl}^{-\beta},
\label{eq3}
\end{equation}

\noindent where $M_{c}$ is the mass that marks the turnover from the power law to the exponential regime. For spiral galaxies $M_{c}$ is found to be $\approx 2\times10^{5}$ M$_{\odot}$ (Jord\'{a}n et al. 2007; Larsen et al. 2009). Eq.~\ref{eq2} represents the general form of an SGMF with a universal IMF. There could be potential dependencies between the mass of the cluster, $M_{cl}$, and the shape of the IMF and/or between $M_{cl}$ and the maximum stellar mass $M_{*,max}$ that can be found in the cluster, as earlier suggested by Weidner \& Kroupa (2004). In our case, we do not assume any relationship between $M_{cl}$ and the shape of the IMF, nor between $M_{cl}$ and $M_{*,max}$ and the analytical description of Eq.~\ref{eq2} can be further simplified by taking $\phi ({\rm log}M_{*})$ out of the integral such that:

\begin{equation}
\Phi\left({\rm log}M_{*},\tau_{cl}\right)=\frac{d{\cal N}}{d{\rm log}M_{*}}=B_{*}\left(\tau_{cl}\right) \phi\left({\rm log}M_{*}\right) \times \int_{M_{cl,min}}^{M_{cl,max}} \xi_{cl} \left(M_{cl}\right) dM_{cl}.
\label{eq4}
\end{equation}

Thus, the term that contains the integration over $M_{cl}$ acts as a simple normalization term and the shape of the SGMF would be indistinguishable from that of the IMF of individual clusters. If the set of parameters that describe the shape of the IMF (i.e., $\theta_{i}$) for the population of young galactic clusters (or cluster that form within a timespan $\tau_{cl}$) each have an intrinsic scatter as suggested by Dib (2014) and Dib et al. (2017), then Eq.~\ref{eq2} should be replaced by

\begin{equation}
\Phi({\rm log}M_{*},\tau_{cl})=B_{*}\left(\tau_{cl}\right) \int_{M_{cl,min}}^{M_{cl,max}}\int_{\theta_{i,min}}^{\theta_{i,max}} \xi_{cl}(M_{cl}) \phi({\rm log}M_{*},\theta_{i}, M_{cl}) P(\theta_{i}) dM_{cl} d\theta_{i},
\label{eq5}
\end{equation}

\noindent where $P(\theta_{i})$ is the probability distribution function of each parameter $\theta_{i}$. When the IMF is described by the TPL (Eq.~\ref{eq1}), Eq.~\ref{eq5} becomes

\begin{equation}
\Phi({\rm log}M_{*},\tau_{cl})=B_{*}\left(\tau_{cl}\right) \int_{M_{cl,min}}^{M_{cl,max}}\int_{\Gamma_{min}}^{\Gamma_{max}} \int_{\gamma_{min}}^{\gamma_{max}} \int_{M_{ch,min}}^{M_{ch,max}} \xi_{cl}(M_{cl}) \phi({\rm log}M_{*},M_{cl},\Gamma,\gamma,M_{ch}) P(\Gamma) P(\gamma) P(M_{ch}) dM_{cl}~d\Gamma~d\gamma~dM_{ch}.
\label{eq6}
\end{equation}

The normalisation coefficient $B_{*}$ is given by the total mass available in all clusters for a given ICMF, $\Sigma_{cl}$, with 

\begin{equation} 
\Sigma_{cl}=B_{*} \left(\tau_{cl}\right) \int_{M_{cl,min}}^{M_{cl,max}} M_{cl}^{-\beta+1} dM_{cl}.
\label{eq7}
\end{equation}

For a given galactic star formation rate ($\rm SFR$) and a timescale of interest ($\tau_{cl}$), the value of $\Sigma_{cl}$ will be given by $\Sigma_{cl} = {\rm SFR} \times \tau_{cl}$, under the assumption that the SFR is constant within the timescale $\tau_{cl}$. In a similar fashion to Eq.~\ref{eq2}, if no correlation between $M_{cl}$ and the shape of the IMF is assumed, as in our case, then Eq.~\ref{eq6} can be further simplified by performing the integration over $M_{cl}$ separately and the integration is reduced to

\begin{equation}
\Phi({\rm log}M_{*},\tau_{cl})=B_{*}\left(\tau_{cl}\right) \int_{M_{cl,min}}^{M_{cl,max}} \xi_{cl}(M_{cl}) dM_{cl} \times \int_{\Gamma_{min}}^{\Gamma_{max}} \int_{\gamma_{min}}^{\gamma_{max}} \int_{M_{ch,min}}^{M_{ch,max}} \phi({\rm log}M_{*},\Gamma,\gamma,M_{ch}) P(\Gamma) P(\gamma) P(M_{ch}) ~d\Gamma~d\gamma~dM_{ch}.
\label{eq8}
\end{equation}

We explore how the shape of the SGMF depends on the level of scatter in the parameters that characterize the IMF of individual clusters. We describe the distributions of $\Gamma$, $\gamma$, and $M_{ch}$ with Gaussian functions:
 
\begin{equation}
P\left(\Gamma \right)=\frac{1}{\sigma_{\Gamma} \sqrt{2 \pi}} \exp\left(-\frac{1}{2} \left(\frac{\Gamma-\bar{\Gamma}}{\sigma_{\Gamma}}\right)^{2}\right),
\label{eq9}
\end{equation}

\begin{equation}
P\left(\gamma \right)=\frac{1}{\sigma_{\gamma} \sqrt{2 \pi}} \exp\left(-\frac{1}{2} \left(\frac{\gamma-\bar{\gamma}}{\sigma_{\gamma}}\right)^{2}\right),
\label{eq10}
\end{equation}

\begin{equation}
P\left(M_{ch} \right)=\frac{1}{\sigma_{M_{ch}} \sqrt{2 \pi}} \exp\left(-\frac{1}{2} \left(\frac{M_{ch}-\bar{M_{ch}}}{\sigma_{M_{ch}}}\right)^{2}\right),
\label{eq11}
\end{equation}

\noindent with mean values  $\bar{\Gamma}=\Gamma_{obs}=1.37$, $\bar{\gamma}=\gamma_{obs}=0.91$, $\bar{M_{ch}}=M_{ch,obs}=0.41$ M$_{\odot}$, which are the values measured for the Milky Way stellar clusters (Dib 2014). The standard deviations of the three parameters are parametrised as  $\sigma_{\Gamma}=a_{\Gamma}  \sigma_{\Gamma_{obs}}$, $\sigma_{\gamma}=a_{\gamma} \sigma_{\gamma,obs}$, and $\sigma_{M_{ch,obs}}=a_{M_{ch}} \sigma_{M_{ch,obs}}$, where $\sigma_{\Gamma_{obs}}$, $\sigma_{\gamma,obs}$, and $ \sigma_{M_{ch,obs}}$ are the observed values by Dib et al. (2014) quoted in \S.~\ref{individual} and where $a_{\Gamma}$, $a_{\gamma}$, and $a_{M_{ch}}$ are free parameters. The work of Dib et al. (2017) suggests that a value of $a_{\Gamma}=a_{\gamma}=a_{M_{ch}}=1$ is related to intrinsic cluster-to-cluster variations in the Milky Way clusters, at least for the sample of 341 clusters they have tested, and that stochastic effects due to random sampling around a universal IMF that can affect low mass clusters are not enough to reproduce the statistics of isolated O stars in Galactic clusters. However, the model presented in this work is agnostic on whether the standard deviations of each of the parameters is due to intrinsic cluster-to-cluster variations or whether there is a contribution from stochastic sampling effects. This is why we consider cases where $a_{\Gamma}$, $a_{\gamma}$, and $a_{M_{ch}}$ are varied in a large range between $0.25$ and $1.5$. 

There is limited information on what the lower and upper limits of these parameters are for Galactic clusters. We adopt here the same limits found by Dib (2014) and used in Dib et al. (2017), namely $\left(\Gamma_{min}=0.7, \Gamma_{max}=2.4\right)$,$\left(\gamma_{min}=0.4,\gamma_{max}=1.5\right)$, and $\left(M_{ch,min}=0.05~{\rm M}_{\odot},M_{ch,max}=1~{\rm M}_{\odot} \right)$. We also assume here that $\beta=2$, $M_{c}=2\times10^{5}$ M$_{\odot}$, and use a value of $\tau_{cl}=12.3$ Myr, and a galactic SFR of $1$ M$_{\odot}$ yr$^{-1}$ (Robitaille \& Whitney 2010). Stellar masses are considered in the range [0.004-120] M$_{\odot}$ and we do not impose any relationship between the cluster mass and maximum stellar mass in the cluster as we have found little evidence for the existence of such a relation (Dib et al. 2017). 

Using the distributions in Eqs.~\ref{eq9}-\ref{eq11}, the integral in Eq.~\ref{eq8} is solved numerically. We first consider the effects on the SGMF of a dispersion in each of the parameters separately. The three stacked plots in the top left panel of Fig.~\ref{fig1} display the effects on the SGMF of the distribution of $\Gamma$ with $P(\Gamma)$ being described by a Gaussian function centred at $\bar{\Gamma}=\Gamma_{obs}=1.37$ and a standard deviation $\sigma_{\Gamma}=a_{\Gamma} \sigma_{\Gamma,obs}$, with $a_{\Gamma}$ that is varied between $0.25$ and $1.5$. The distributions of $\gamma$ and $M_{ch}$ ($P(\gamma)$ and $P(M_{ch})$) are taken to be delta functions located at the positions of their observed values by Dib (2014), namely $\gamma_{obs}=0.91$, and M$_{ch,obs}=0.41$ M$_{\odot}$. A larger cluster-to-cluster scatter in $\Gamma$ results in an increased deviation from a pure power-law function at the high mass end and to a shallower slope of the SGMF. The same effect has been observed by Cervi\~{n}o \& Luridiana (2006) who explored the existence of scatter in the high-mass slope on a single power-law function. The three sub panels compare our results to the Galactic stellar mass function inferred for the nearby Galactic field by Kroupa (2001), Chabrier (2005), and Parravano et al. (2011). Ideally, the results of our models should be compared to a single generation of stars in a galaxy. In the Milky Way, the best comparison could be performed with the young stellar populations (i.e., stars with ages $\lesssim 10-12$ Myrs). However, the mass function of young stars that are present both in clusters and in the field of the Milky Way is not well established. It should be noted that the slope of the Galactic field stellar mass function adopted by Chabrier (2005) and Parravano et al. (2011) is the one derived by Salpeter (i.e., $\Gamma=1.35$), while the value adopted by Kroupa (also $\Gamma=1.35$) is a mean value derived for Galactic clusters in the Milky Way. These values are not corrected for the effects of stellar evolution. Our results yield a shallower slope than the Galactic field in the high stellar mass regime (due to contributions from the wings in $P(\Gamma)$), and also do not take into account corrections for the effects of stellar evolution. While this is beyond the scope of the present work, we anticipate that taking into account such corrections in our models would steepen the slope of the SGMF in this mass range and improve the fit to the Galactic field stellar mass function.

Similarly, the top right panel and bottom left panels in Fig.~\ref{eq1} display the effects on the SGMF of scatter in $\gamma$ and in M$_{ch}$. The dispersion in theses case is described by Gaussian functions centred at their observed values and the standard deviations are varied such that $a_{\gamma}$ and $a_{M_{ch}}$ are varied between 0.25 and 1.5. The distributions of the parameters that are not varied are described by delta functions located at their observed positions. As for the case with $\Gamma$, a larger dispersion in the parameter $\gamma$ leads to more flattening of the slope of the SGMF in this mass range. However, the effects of even the largest dispersion considered for this parameter (i.e., $a_{\gamma}=1.5$) do not lead to a substantial flattening of the low mass slope of the SGMF. A $1-\sigma$ dispersion in $M_{ch}$ of the order of the one measured by Dib (2014) (i.e., $\approx 0.3$ M$_{\odot}$), leads to two interesting features in the SGMF. Firstly, cluster-to-cluster variations of $M_{ch}$ lead to the formation of a "plateau" in the SGMF, similar to the one suggested in the Kroupa Galactic mass function. Secondly, this level of dispersion in $M_{ch}$ leads to a flattening of the SGMF in the low mass end, even if the mean value of $\gamma$ for the ensemble of clusters is much steeper ($\bar{\gamma}=0.91$).

The bottom right panel in Fig.~\ref{fig1} displays cases where all the three parameters have a scatter and their distributions functions are described by Gaussian functions whose mean values are fixed at the observed values and standard deviations are varied between 0.25 and 1.5 times the observed values. When all three parameters have cluster-to-cluster variations, all the features in which they influence the SGMF individually are preserved, namely a shallower slope and a deviation from a pure power-law function in the intermediate- to high stellar mass regime, a shallower slope than that of individual clusters in the low stellar mass regime, and the formation of a plateau in the low- to intermediate stellar mass regime. The slope of the SGMF in the low mass regime (i.e., $M_{*} < 0.03$ M$_{\odot}$) is $0.89$, $0.85$, $0.81$, $0.78$, and $0.75$ for the cases were $a_{\Gamma}=a_{\gamma}=a_{M_{ch}}=0.25,0.5,0.7,1$, and $1.5$, respectively, which are shallower than the mean value of 0.91 for observed clusters. It is interesting to note that a model with $(a_{\Gamma},a_{\gamma},a_{M_{ch}})$=$(0.75,0.75,0.5)$ provides an excellent fit to the Kroupa Galactic field mass function and is also compatible with the Parravano et al. (2011) Galactic field mass function. Further work is needed to investigate what would be the effect of more complex parameter distributions of the IMF parameters when more constraints from observations become available. The sample of clusters from the MWSC survey used by Dib et al (2017) to constrain the distributions of the IMF parameters did not distinguish between bound clusters and more loose stellar associations. Recently, Chandar et al. (2017) argued that the fraction of young stars in compact clusters in nearby galaxies is $\approx 25\%$ of the total young stellar population. It would be interesting to explore whether the IMF parameters for hierarchical structures have distinct distributions of the IMF parameters than the ones of more compact bound clusters. Would that turn out to be the case, it would be necessary to adopt more complex distributions (e.g., bi-modal distributions) of the IMF parameters in order to construct the SGMF. With the current observational constraints, the main result of this paper is that a SGMF which resembles the Galactic field mass function can result from the summation of a large population of stellar clusters which have a significant level of variations among their IMFs, and which can be, individually, very different from the SGMF.  

\begin{acknowledgements}
We thank the anonymous referee for useful comments and suggestions. SD acknowledges the hospitality of the University of Western Ontario where part of this work has been completed. This research has made use of NASA's Astrophysics Data System Bibliographic Services. SB was supported by a Discovery grant from NSERC.
\end{acknowledgements}

%

\begin{thebibliography}{}
\bibitem[2004] {basu04} Basu, S., Jones, C. E. 2004, MNRAS, 347, 47
\bibitem[Basu (2015)] {basu15} Basu, S., Gil, M., Auddy, S. 2015, MNRAS, 449, 2413
\bibitem[Bekki (2014)] {bekki14} Bekki, K., Tsujimoto, T. 2014, MNRAS, 444, 3879
\bibitem[Bochanski (2010)] {bochanski10} Bochanski, J. J., Hawley, S. L., Covey, K. R., West, A. A., Reid, I. N., Golimowski, D. A., Ivezi\'{c}, Z. 2010, AJ, 139, 2679
\bibitem[Cervino (2006)] {cervino06} Cervi\~{n}o, M., Luridiana, V. 2006, A\&A, 451, 475
\bibitem[Chabrier  (2005)] {chabrier05} Chabrier, G. 2005, in Astrophysics and Space Science Library, Vol. 327, The Initial Mass Function 50 Years Later, ed. E. Corbelli, F. Palla, \& H. Zinnecker, 41
\bibitem[Chandar (2017)] {chandar17} Chandar, R., Fall, S. M., Whitmore, B. C., Mulia, A. J. 2017, ApJ, 849, 128
\bibitem[De Marchi (2010)] {demarchi10} De Marchi, G., Paresce, F., Portegies Zwart, S. 2010, ApJ, 718, 23
\bibitem[Dib (2006)] {dib06} Dib, S., Bell, E., Burkert, A. 2006, ApJ, 638, 797
\bibitem[Dib (2007)] {dib07} Dib, S., Kim, J., Shadmehri, M. 2007, MNRAS, 381, L40
\bibitem[Dib (2008)] {dib08} Dib, S., Shadmehri, M., Gopinathan, M., Kim, J., Henning, T. 2008, in Beuther, H., Linz, H., Henning, T., eds, ASP Conf. Ser. Vol 387, Massive Star Formation: Observations Confront Theory. Astron. Soc. Pac., San Francisco, p. 282
\bibitem[Dib (2010)] {dib10} Dib, S., Shadmehri, M., Padoan, P., Maheswar, G., Ojha, D. K., Khajenabi, F. 2010, MNRAS, 405, 401
\bibitem[Dib (2011)] {dib11a} Dib, S., Piau, L., Mohanty, S., Braine, J. 2011, MNRAS, 415, 3439
\bibitem[Dib (2011)] {dib11b} Dib, S. 2011, ApJ, 737, L20
\bibitem[Dib (2013)] {dib13} Dib, S., Gutkin, J., Brandner, W., Basu, S. 2013, MNRAS, 436, 3727
\bibitem[Dib (2014)] {dib14} Dib, S. 2014, MNRAS, 444, 1957 
\bibitem[Dib (2017)] {dib17} Dib, S., Schmeja, S., Hony, S. 2017, MNRAS, 464, 1738
\bibitem[Haas (2010)] {haas10} Haas, M. R., Anders, P. 2010, A\&A, 512, 79 
\bibitem[Haugbolle (2017)] {haugboelle17} Haugb{\o}lle, T., Padoan, P., Nordlund, \AA, 2017, submitted, arXiv:1709.01078
\bibitem[Hony (2015)] {hony15} Hony, S., Gouliermis, D. A., Galliano, F., Galametz, M., Cormier, D., Chen, C.-H. R., Dib, S. et al. 2015, MNRAS, 448, 1847
\bibitem[Jordan (2007)] {jordan07} Jord\'{a}n, A., McLaughlin, D. E., C\^{o}t\'{e}, P. et al. 2007, ApJS, 171, 101 
\bibitem[Kharchenko (2013)] {kharchenko13} Kharchenko, N. V., Piskunov, A. E., Schilbach, E., R\"{o}ser, S., Scholz, R.-D. 2013, A\&A, 558, 53
\bibitem[Kroupa (2001)] {kroupa01} Kroupa, P. 2001, MNRAS, 322, 231
\bibitem[Kroupa (2013)] {kroupa13} Kroupa, P., Weidner, C., Pflamm-Altenburg et al. 2013, Planets, Stars and Stellar Systems. Volume 5: Galactic Structure and Stellar Populations, 5, 115
\bibitem[Larsen (2009)] {larsen09} Larsen, S. S. 2009, A\&A, 494, 539 
\bibitem[Mallick (2014)] {mallick14} Mallick, K. K., Ojha, D. K., Tamura, M., Pandey, A. K., Dib, S. et al. 2014, MNRAS, 443, 3218
\bibitem[Martizzi (2016)] {martizzi16} Martizzi, D., Fielding, D., Faucher-Gigu\`{e}re, C.-A., Quataert, E. 2016, MNRAS, 459, 2311
\bibitem[Myers (2009)] {myers09} Myers, P. C. 2009, ApJ, 706, 1341
\bibitem[Nakamura (2007)] {nakamura17} Nakamura, F, Li, Z.-Y. 2007, ApJ, 662, 395
\bibitem[Padoan (2016)] {padoan16} Padoan, P., Pan, L., Haugb{\o}lle, T., Nordlund, \AA.  2016, ApJ, 822, 11
\bibitem[Parravano (2011)] {parravano11} Parravano, A., McKee, C. F., Hollenbach, D. J. 2011, ApJ, 726, 27
\bibitem[Robitaille (2010)] {robitaille10} Robitaille, T. P., Whitney, B. A. 2010, ApJ, 710, 11
\bibitem[Rybizki (2015)] {rybizki15} Rybizki, J., Just, A. 2015, MNRAS, 447, 3880
\bibitem[Salpeter (1955)] {salpeter55} Salpeter, E. E. 1955, ApJ, 121, 161
\bibitem[Scalo (1986)] {scalo86} Scalo, J. M. 1986, Fundam. Cosm. Phys., 11,1
\bibitem[Schmeja (2014)] {schmeja14} Schmeja, S., Kharchenko, N. V., Piskunov, A. E. et al. 2014, A\&A, 568, 51
\bibitem[Shadmehri (2004)] {shadmehri04} Shadmehri, M. 2004, MNRAS, 354, 375
\bibitem[Weidner (2004)] {weidner04} Weidner, C., Kroupa, P. 2004, MNRAS, 348, 187
\bibitem[Weidner (2005)] {weidner05} Weidner, C., Kroupa, P. 2005, ApJ, 625, 754
\bibitem[Weidner (2013)] {weidner13} Weidner, C., Kroupa, P., Pfamm-Altenburg, J., Vazdekis, A. 2013, MNRAS, 436, 3309
\bibitem[Zaritsky (2014)] {zaritsky14} Zaritsky, D., Colucci, J. E., Pessev, P. M. et al. 2014, ApJ, 796, 71
\end{thebibliography}
%

\end{document}